\documentclass[]{mn2e}
\newcommand{\be}{\begin{eqnarray}}
\newcommand{\ee}{\end{eqnarray}}

\newcommand{\msun}{{\rm M}_{\odot}}

\newcommand{\me}{{\rm M}_{\oplus}}

\newcommand{\gcm}{ \; {\rm g} \; {\rm cm}^{-2}}
\newcommand{\tm}{ t_{\rm mig}}
\newcommand{\te}{ t_{\rm ecc}}
\newcommand{\tg}{ t_{\rm gap}}
\newcommand{\tnu}{ t_{\nu}}
\newcommand{\tst}{ t_{\rm start}}

\newcommand{\ls}{\raisebox{-.8ex}{$\buildrel{\textstyle<}\over\sim$}}
\usepackage{graphicx}
\usepackage{amsmath}
\usepackage{caption}
\usepackage{float}
\usepackage{color}
\graphicspath{{figures/}}

\begin{document}

\title[Planetary systems in transition discs]{On the formation of
  planetary systems in photoevaporating transition discs}

\author[Caroline Terquem] { Caroline Terquem \\ Physics Department,
  University of Oxford, Keble Road, Oxford OX1 3RH, UK \\
  Institut d'Astrophysique de Paris, UPMC Univ Paris 06, CNRS,
  UMR7095, 98 bis bd Arago, F-75014, Paris, France \\ E-mail:
  caroline.terquem@physics.ox.ac.uk }

\date{}

\pagerange{\pageref{firstpage}--\pageref{lastpage}} \pubyear{}

\maketitle

\label{firstpage}

%
%===========================================================
%

\begin{abstract}
% 250 words max
  In protoplanetary discs, planetary cores must be at least 0.1~$\me$
  at 1 au for migration to be significant; this mass rises to 1~$\me$
  at 5~au.  Planet formation models indicate that these cores form on
  million year timescales.  We report here a study of the evolution of
  0.1~$\me$ and 1~$\me$ cores, migrating from about 2 and 5~au
  respectively, in million year old photoevaporating discs.  In such a
  disc, a gap opens up at around 2~au after a few million years. The
  inner region subsequently accrete onto the star on a smaller
  timescale.  We find that, typically, the smallest cores form systems
  of non--resonant planets beyond 0.5~au with masses up to about
  1.5~$\me$. In low mass discs, the same cores may evolve {\em in
    situ}.  More massive cores form systems of a few earth masses
  planets.  They migrate within the inner edge of the disc gap only in
  the most massive discs.  Delivery of material to the inner parts of
  the disc ceases with opening of the gap.  Interestingly, when the
  heavy cores do not migrate significantly, the type of systems that
  are produced resembles our solar system.  This study suggests that
  low mm flux transition discs may not form systems of planets on
  short orbits but may instead harbour earth mass planets in the
  habitable zone.
  
\end{abstract}


%
%===========================================================
%

\section{Introduction}
\label{sec:intro}

Before the discovery of extrasolar planets, explaining the formation
of terrestrial planets in our solar system was already a challenging
task.  The detection of a large variety of planetary systems
containing Earth--mass or super--Earth planets by the {\em Kepler}
satellite has made the task even more arduous.  Terrestrial mass
planets are usually considered to form either {\em in situ} or through
inward migration.  {\em In situ } formation was the early scenario
proposed for forming planets in our solar system (Wetherill 1988,
Lissauer 1993 and references therein), and has been put forward as a
way to explain the {\em Kepler} candidates (Hansen \& Murray 2013).
However, it has been argued that this model is not consistent with the
distribution of solids in discs (Raymond \& Cossou~2014,
Schlichting~2014).  Also, {\em in situ} formation models have
difficulties forming cores of giant planets before the gas dissipates
(Thommes et al. 2003, Chambers 2016).  Formation through inward
migration is an efficient way of obtaining tight systems of
super--Earths with short periods (Terquem \& Papaloizou 2007,
Haghighipour 2013 and references therein), but fails to explain
terrestrial planets similar to those in our solar system.  However,
migration of planets in dissipating discs has recently been studied
(Coleman \& Nelson 2014, Cossou et al. 2014, Coleman
  \& Nelson 2016) and seems to offer a way of combining the
advantages of both models for forming terrestrial planets.

Coleman \& Nelson (2014, 2016) studied the evolution
  of systems containing initially 36 or 52 cores with masses of $0.3$
  or $0.1 \; \me$, respectively, spread between 1 and 20~au.  In
  addition, there were thousands of planetesimals with masses 10, 20
  or 50 times smaller than that of the embryos and distributed in
  between the planets.  The disc had a surface gas density at least
equal to that of the minimum solar mass nebula, i.e.
$\Sigma \propto r^{-3/2}$ and $\Sigma=1.7 \times 10^3 \; \gcm$ at
1~au, and was subject to photoevaporation and viscous evolution.  They
found that terrestrial--mass planets and super--Earths formed in the
discs with the lowest masses.  In the models with
  small abundance of solids and large planetesimals, growth was found
  to be limited so that migration was inefficient.  Such models
  resulted in systems of low mass planets spread out through the disc
  and in which mean motion resonances were destroyed after the disc
  dissipated.

Cossou et al. (2014) started with cores with masses between 0.1 and
2~$\me$ spread between 1 and 20~au. The total mass in the cores was
between 21 and 84~$\me$.  They considered a disc with a surface gas
density of $300 \; \gcm$ at 1~au and $\Sigma \propto r^{-1/2}$.  In
some of their simulations, the disc's mass was decreased exponentially
to mimic disc's viscous evolution and photoevaporation.  Systems of
hot super--Earths in mean motion resonances were produced in discs
which were not dissipating, whereas the systems were more spread out
and not in resonances when dissipation was included.

In these studies, all types of planets formed from the same parent
population and, in general, the planetary systems that were produced
at around 1~au contained planets more massive than the terrestrial
planets in our solar system.  In this paper, we consider a model where
different parent populations exist at different locations and the disc
is photoevaporating.
 
%Observations suggest that the surface density in discs varies like
%$r^{-1}$, and that the mass within about $ 50$~au in one~Myr--old
%discs is on the order of 0.01~$\msun$ (Andrews et al. 2010).  
In typical protoplanetary discs, the type--I migration timescale
becomes comparable to the disc lifetime for cores at least as massive
as 0.1 and 1~$\me$ at 1 and 5~au, respectively.  Planet formation
models suggest that it takes at least 1~Myr for 0.1~$\me$ cores to
form at $\sim 1$~au.  More massive cores, with a mass $\sim 1 \; \me$,
may form on this timescale beyond the snow line, at around 5~au.  This
prompts us to consider a model where the initial conditions are a
population of 0.1~$\me$ cores between 1 and 5~au and a population of
1~$\me$ cores beyond 5~au, which start migrating when the disc is
$\sim 1$~Myr--old.
% at which point the surface density at 1~au is
%between a few hundred and $10^3 \; \gcm$.  
The total mass in these populations of cores is set by the initial
surface density in the disc.

We evolve these initial populations in a disc which undergoes a
transition due to X--ray photoevaporation.  Transition discs are
defined as discs lacking emission in the near--infrared, which means
that there is a (large) hole in the dust distribution in their inner
parts.  Observations strongly suggest two types of
transition discs: those with low mm flux, which have low accretion
rates and hole sizes smaller than about 20~au, and those with high mm
flux, which have higher accretion rates and hole sizes larger than
20~au (Owen \& Clarke 2012, Owen 2016 and references therein).  It is
believed that low mm flux transition discs are in the process of
dispersing, whereas high mm flux transition discs are not.  The most
commonly accepted interpretation for low mm flux transition discs is
X--ray photoevaporation.  In this model, a gap opens up after about
3~Myr (75\% of the disc lifetime) at $\sim 2$~au, where the accretion
rate in the disc matches the photoevaporation rate (Owen, Ercolano \&
Clarke 2011).  The inner parts of the disc then become decoupled from
the outer parts and cannot be resupplied in gas and dust.  They
subsequently accrete onto the central star whereas the outer edge of
the gap recedes due to photoevaporation.  By contrast, it is believed
that, in high mm flux transition discs, a massive planet (with a mass of
a few Jupiter masses) is responsible for creating a gap.

The above discussion suggests that, in low mm flux transition discs,
0.1~$\me$ and 1~$\me$ cores forming between 1 and 5~au and beyond
5~au, respectively, on a timescale of 1~Myr, would start to migrate a
million years or so before a gap opens up at around 2~au.  As their
migration timescale itself is on the order of a million years, it can
be expected that the dispersion of the disc will prevent migration of
cores to very small radii and also that the formation of the gap will
prevent massive cores to be delivered to the region of terrestrial
planets.  This is the model we explore in this paper.

The plan of the paper is as follows.  In section~\ref{sec:formation},
we describe the disc evolution model and justify the parameters that
are used in the simulations.  We also give expressions for the planet
migration and eccentricity damping timescales and briefly discuss core
formation timescales.  In section~\ref{sec:sim}, we present $N$--body
simulations of cores migrating in photoevaporating discs.  We first
describe the numerical scheme and the initial set up.  We then present
the results of the simulations.  We show that, for reasonable
parameters that are consistent with the observations, a population of
$0.1 \; \me$ cores originating from between 2 and 4~au migrate down to
0.5--1~au.  The final masses of planets are between a fraction of an
Earth mass and $\sim 1.5 \; \me$.  As for the population of 1~$\me$
cores originating from $\sim 5$~au, it forms a few cores of a few
Earth masses which may migrate down below the inner edge of the gap
only in the most massive discs considered here.  In less massive
discs, cores with a mass comparable to that of Jupiter may be left at
a few au from the star.  Finally, in section~\ref{sec:discussion}, we
summarize and discuss our results.

%
%===========================================================
%

\section{Disc evolution and planet formation}
\label{sec:formation}

In this section, we review the different timescales that are used in
the simulations. 

%{\bf Core of 0.1 earth mass form in 1 Myr when Sigma is down to 100 g/cm2 --> they start to migrate
% on a timescale 10^7 years; the lifetime of the disc from that point
% on is on the order of Myr, and therefore the cores can never reach
%  the inner edge.  Furthermore, dissipation of the disc on a timescale
%  on the order of 0.1 Myr prevents MMR.}

%{\bf justifiy that the timescale to form 0.1 earth mass core is about
%  1 Myr.  Chambers proposes a mechanism to go faster but he has no
%  migration, so he can't form planets otherwise...}

\subsection{Disc evolution and parameters}

\label{sec:disc}

We adopt an initial surface density profile in the disc:

\begin{equation}
\Sigma(r,t=0) = \Sigma_1  \left( \frac{r}{1 \; {\rm au}} \right)^{-1},
\label{sigma}
\end{equation}

\noindent where $\Sigma_1$ is the initial surface density at 1~au.  This
density gradient gives the best fit to the thermal continuum emission
from $\sim 1$~Myr old discs in the Ophiuchus star--forming region, as
shown by Andrews et al. (2010).  We express $\Sigma_1$ in terms
of the disc mass within 50~au, $M_{50}$:

\begin{equation}
\Sigma_1 = 2.8 \times 10^4 \; \frac{M_{50}}{ \msun} \gcm.
\label{sigma1}
\end{equation}

In this paper, we focus on the evolution of planetary systems in
transition discs which have low mm fluxes and are supposed to be the
consequence of X--ray photoevaporation (Owen~2016 and references
therein).  In this model, a gap first opens up in the disc at the
location where the accretion rate becomes equal to the
photoevaporation rate.  At this point, the parts of the disc within
the inner edge of the gap become isolated from the outer parts, and
cannot be resupplied in gas and dust. They accrete onto the central
star on a timescale of a few $10^5$~years and become depleted.  At the
same time, the outer edge of the gap moves out because of erosion due
to photoevaporation.  The evolution of a disc subject to such a
process has been calculated by Owen, Ercolano \& Clarke (2011).  Here,
we evolve the surface density profile $\Sigma(r,t)$ in such a way as
to reproduce their calculation.

We start with $\Sigma(r,0)$ given by equation~(\ref{sigma}).  After a
time $t= \tg$ which is between 1 and 3 Myr ($\sim$ 75\% of the disc
lifetime, see Owen et al. 2011), a gap opens up between the radii that
we fix to be $r_{10} = 2$~au and $r_{20} = 3$~au.  Between $t=0$ and
$ \tg $, we assume that $\Sigma$ decreases linearly with time, i.e.:

\begin{equation}
\Sigma(r,t)=\Sigma(r,0) \left[ 1- \frac{(1-\eta)t}{ \tg } \right],
\; \; {\rm for} \; \; 0 \le t \le \tg ,
\label{sigmatgap}
\end{equation}

\noindent where $\eta \equiv \Sigma(r,
\tg)/ \Sigma(r,0) $ is the fraction of mass left after a time
$\tg$.
Subsequently, the inner disc disperses on a timescale $\tnu$
such that: 

\begin{equation}
\begin{split}
  \Sigma(r_{\rm cav} \le r \le r_{10},t)  =\eta \Sigma(r,0) \left[ 1- \frac{(t-\tg)}{ \tnu } \right]
                                           ,  \; \; \\ 
{\rm for} \; \;
                                           \tg \le t \le \tg + \tnu, 
\end{split}
\end{equation}

\begin{equation}
  \Sigma(r \le r_{10} ,t)  =0 ,  \; \; {\rm for} 
                            \; \; t \ge \tg + \tnu,
\end{equation}

\noindent where $r_{\rm cav}=0.05$~au if the radius of the inner disc
cavity produced by the magnetic interaction between the disc and the
star.  We will take $\tnu \sim 10^4$--$10^5$~yr as this is the viscous
timescale at 2~au in a standard $\alpha$--disc model (Shakura \&
Sunyaev~1973) with $\alpha$ between $10^{-3}$ and $0.01$.  After the
gap opens up, its outer edge $r_2$ moves out according to:

\begin{equation}
r_2=r_{20}+(10 \; {\rm au}-r_{20}) \; \frac{t-\tg}{3.2 \times 10^6 \;
  {\rm yr} - \tg} ,  \; \; {\rm for} \; \;  t \ge \tg , 
\end{equation}

\noindent so that after $3.2 \times 10^6$~yr ($\sim$ 80\% of the disc
lifetime, see Owen et al. 2011), the outer edge has moved up to 10~au.
The surface density does not vary significantly beyond $r_2$:

\begin{equation}
\begin{split}
\Sigma(r \ge r_{2},t) = \Sigma(r \ge r_{2}, \tg ) = \eta \Sigma(r \ge
r_{2}, 0 ), \; \;  \\
 {\rm for} \; \;  t \ge \tg ,
\end{split}
\end{equation}

\noindent while it is zero inside the gap:

\begin{equation}
\Sigma(r_{10} \le r \le r_{2},t) = 0, \; \; {\rm for} \; \;  t \ge \tg.
\end{equation}

In the numerical simulations presented below, we will take
$\Sigma_1 \sim 10^3 \gcm$ at $t=0$, i.e.
$M_{50} \; \ls \; 0.1$~$\msun$ initially.  More massive discs would be
gravitationally unstable.  It has been pointed out that the observed
range of accretion rates in T~Tauri stars require an initial disc mass
between 0.01 and 0.2~$\msun$ (Hartmann et al.~1998).  Masses derived
for the $\sim 1$~Myr old discs in the Ophiuchus star--forming region
are on the order of $0.01$~$\msun$ within 50~au (Andrews et
al.~2010).  This suggests that $\eta=0.1$ in
equation~(\ref{sigmatgap}).  A decrease in disc's mass by a factor 10
during the first Myr of evolution is also consistent with the
observations and modelling of discs around T~Tauri stars in the Taurus
and Chamaelon~I molecular clouds complex (Hartmann et al. 1998).

\begin{figure}
\begin{center}
\includegraphics[scale=0.45]{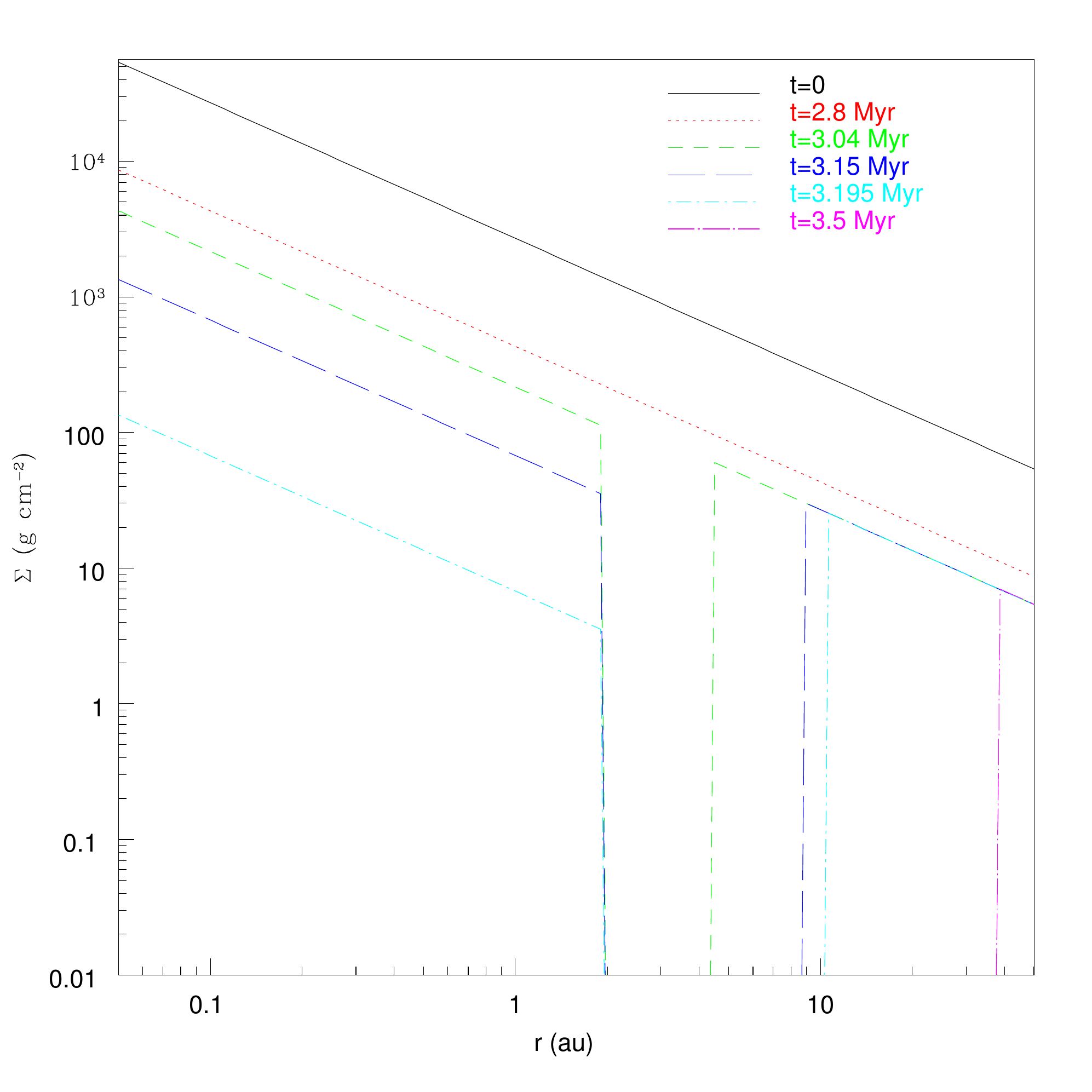}
\end{center}
%\vspace{-4.cm}
\caption{Surface density $\Sigma$ (in $\gcm$) {\em versus} $r$ (in au)
  in logarithmic scale for $\tg=3$~Myr, $\tnu=0.2$~Myr and
  $\Sigma_1=2.7 \times 10^3 \gcm$.  The different lines correspond to
  the following times: $t=0$ ({\em solid line}), $t=2.8$~Myr ({\em
    dotted line}), $t=3.04$~Myr ({\em short--dashed line}),
  $t=3.15$~Myr ({\em long--dashed line}), $t=3.195$~Myr ({\em
    dotted--short--dashed line}) and $t=3.5$~Myr ({\em
    dotted--long--dashed line}).  This figure is similar to fig.~(9)
  of Owen et al. (2011). }
\label{fig1}
\end{figure}

The surface density at different times is shown in figure~(\ref{fig1})
for $\tg=3$~Myr, $\tnu=0.2$~Myr and $\Sigma_1=2.7 \times 10^3 \gcm$.

With the initial surface density of gas given by
equation~(\ref{sigma}), and adopting a mass dust--to--gas ratio of
0.01, we calculate that the initial mass of dust in an annulus between
two radii $r_{\rm in}$ and $r_{\rm out}$ is:

\begin{align}
M_{\rm dust} & = 2 \pi ( 1 \; {\rm au})^2 \Sigma_1 \left( \frac{r_{\rm
      out}}{1 \; {\rm au}} - \frac{r_{\rm
      in}}{1 \; {\rm au}} \right) \nonumber \\
& \simeq 2.4~\me \left(  \frac{\Sigma_1}{10^{3}
\; \gcm}   \right) \left( \frac{r_{\rm
      out}}{1 \; {\rm au}} - \frac{r_{\rm
      in}}{1 \; {\rm au}} \right) .
\label{Mdust}
\end{align}

\subsection{Migration and eccentricity damping timescales}

Tidal interaction between a planet and the disc in which it is
embedded leads to a change of the planet's angular momentum (i.e.
migration) and to eccentricity damping on the timescales $\tm$ and
$\te$, respectively.  The cores we consider here are small enough that
they undergo inwards type~I migration.  In this regime, we use for
$\tm$ the timescale derived by Tanaka et al. (2002) and modified by
Papaloizou \& Larwood (2000) to account for the effect of a finite
eccentricity:

\begin{equation}
\begin{split}
\tm ({\rm yr}) = \frac{2}{2.7+1.1 n}  \; 
\left( \frac{\msun}{m_p}  \right)
\left( \frac{H}{r} \right)^2
\frac{\msun}{2 \pi \Sigma (1 \; {\rm au})^2 }
 \\
\times \left( \frac{1 \; {\rm au}}{a} \right)^{1/2} 
\left[ 1+ \left( \frac{ e }{1.3 H/r} \right)^5 \right]
\left[ 1- \left( \frac{ e }{1.1 H/r}\right)^4 \right]^{-1} ,
\label{tmig}
\end{split}
\end{equation}

\noindent where $m_p$, $a$ and $e$ are the mass, semimajor axis and
eccentricity of the planet, $\Sigma$ is the disc  surface density of
gas and $H/r$ is the disc aspect ratio at the location of
the planet, and $n \equiv - d \ln \Sigma / d \ln r$.  Here we have
assumed that the central object is a solar mass star.  Thereafter, we will take
$n=1$, as in section~\ref{sec:disc} (see eq.~[\ref{sigma}]).

\noindent For $\te$, we adopt the timescale given by Tanaka \& Ward
(2004) and again modified by  Papaloizou \& Larwood (2000) to account
for the effect of the eccentricity:

\begin{equation}
\begin{split}
 \te  ({\rm yr}) = 0.1 \left( \frac{\msun}{m_p}  \right)
\left( \frac{H}{r} \right)^4
\frac{\msun}{2 \pi \Sigma (1 \; {\rm au})^2 }
\left( \frac{1 \; {\rm au}}{a} \right)^{1/2}  \\
\times \left[ 1+ 0.25 \left( \frac{ e }{H/r}\right)^3 \right],
\label{tecc}
\end{split}
\end{equation}

\noindent where the factor of $0.1$ is taken from Cresswell \& Nelson
(2006) as it gives good agreement between equation~(\ref{tecc}) and
the damping timescale obtained in hydrodynamical simulations.

For a planet with $e=0$ at $a=1$~au in a disc with $n=1$ and
$H/r=0.05$, equations~(\ref{tmig}) and~(\ref{tecc}) can be written as:

\begin{align}
 \tm ({\rm yr} )  & \simeq  6
\times 10^5 \; \left[  \frac{\Sigma({\rm 1 \; au})}{10^{3}
\; \gcm}   \right]^{-1} \;
\left( \frac{ m_p}{  \me} \right)^{-1}  \label{tmig2} \\
\te ({\rm yr} )  & \simeq  300 \; \left[  \frac{\Sigma({\rm 1 \; au})}{10^{3}
\; \gcm}   \right]^{-1}  \; \left( \frac{ m_p }{\me } \right)^{-1},
\end{align}

\noindent where $\Sigma({\rm 1 \; au})$ means that $\Sigma$ is
evaluated at 1~au.

Note that radiation--hydrodynamical simulations of disc/planet
interactions have shown that the corotation torque could lead to
outwards migration, but that does only affect cores with masses
between about 4 and 30~$\me$ (and eccentricities below $\sim 0.015$),
larger than the cores we consider here (Paardekooper \& Mellema 2006,
Kley, Bitsch \& Klahr 2009, Bitsch \& Kley 2010).

\subsection{Planet formation timescale}

\label{sec:planet}

Assuming that the surface density of gas at 1~au is smaller than
about $10^3 \; \gcm$ (see section~\ref{sec:disc}), equation~(\ref{tmig2})
indicates that migration at $\sim 1$~au only affects cores that have
a mass larger than about 0.1~$\me$ (Mars's mass).  Smaller mass
objects would migrate on a timescale longer than the disc's lifetime,
which is of a few~Myr.

Cores with masses $\sim 0.1 \; \me$ form through oligarchic growth
on a timescale which, at $\sim 1$~au, is on the order of at least
1~Myr (see Chambers 2016 and references therein and Kobayashi \&
Dauphas~2013 for the formation of Mars).  Therefore, migration starts
to be significant when the disc's mass is about one tenth of its
initial value, i.e. $M_{50} \sim 0.01 \; \msun$ and
$\Sigma({\rm 1 \; au})  \sim 10^2 \; \gcm$.

Beyond the snow line, it is expected that growth is faster, although
there is still much uncertainty about building massive cores on a
short timescale there (Chambers 2016).  In the simulations below, we
will consider cases where 1~$\me$ cores have formed on a timescale of
1~Myr at around 5~au.

%-----------------

\section{Numerical simulations}
\label{sec:sim}

%-----------------

In this section, we study the evolution of a population of small cores
originating from the region of the terrestrial planets, that of a
population of earth mass cores originating from beyond the snow line,
and finally the evolution of a mixture of both small and large cores.

\subsection{Numerical integration}

To compute the evolution of a population of cores migrating through a
disc, we use the $N$--body code described in Papaloizou~\&
Terquem~(2001) in which we have added the effect of the disc torques
(see also Terquem \& Papaloizou 2007).  The equations
  of motion are integrated using the Bulirsch--Stoer method with a
  timestep which is adjusted to match a prescribed accuracy (e.g.,
  Press et al. 1992).

The equations of motion for each core are:

\begin{equation} {{\rm d}^2 {\boldsymbol{r}}_i\over {\rm d}t^2} = -{G \msun
    \boldsymbol{r}_i \over |\boldsymbol{r}_i|^3} -\sum_{j=1 \ne i}^N
  {G m_j \left(\boldsymbol{r}_i- \boldsymbol{r}_j \right) \over 
     |\boldsymbol{r}_i- \boldsymbol{r}_j |^3} - \sum_{j=1}^N 
     {G m_j \boldsymbol{r}_{j} 
     \over |\boldsymbol{r}_{j}|^3} +
     \boldsymbol{\Gamma}_{i}  \; ,
\label{emot}
\end{equation} 

\noindent where $G$ is the gravitational constant and $m_i$
and $\boldsymbol{r}_i$ denote the mass  
and  position vector of core $i$, respectively.  The
third term on the right--hand side is the acceleration of the
coordinate system based on the central star (indirect term).

\noindent Acceleration due to tidal interaction with the disc is dealt
with through the addition of extra forces as in Papaloizou \& Larwood
(2000, see also Terquem \& Papaloizou 2007):

\begin{equation}
\boldsymbol{\Gamma}_{i} = 
-\frac{1}{\tm} \frac{{\rm d} \boldsymbol{r}_i}{{\rm d}t} -
\frac{2}{| \boldsymbol{r}_i|^2 \te} 
\left( \frac{{\rm d}  \boldsymbol{r}_i}{{\rm d}t} \cdot
\boldsymbol{r}_i \right) 
\boldsymbol{r}_i ,
\end{equation}

\noindent where $\tm$ and $\te$ are given by equations~(\ref{tmig})
and~(\ref{tecc}) in which $m_p$ is replaced by $m_i$.  Note that the
timescale on which the semimajor axis decreases is $\tm/2$ (e.g.,
Teyssandier \& Terquem 2014).  As here cores never approach the star
very closely, we do not include contribution from the tides raised by
the star nor from relativistic effects.

Collisions between cores are dealt with in the following way: if the
distance between cores~$i$ and~$j$ becomes less than $R_i+R_j$, where
$R_i$ and $R_j$ are the radii of the cores, a collision occurs and the
cores are assumed to merge.  They are subsequently replaced by a
single core of mass $M_i+M_j$ with the position and the velocity of
the center of mass of cores~$i$ and~$j$.

\subsection{Initial set up}

We consider a disc which has either $\Sigma_1=2.7 \times 10^3 \; \gcm$
or $900 \gcm$, i.e. $M_{50}=0.1$ or $0.03 \; \msun$ initially.  Note
that this latter value is about twice as large as the minimum mass
solar nebula (Hayashi 1981).  We start with a population of $N$
cores on circular orbits in the disc midplane spread between an inner
radius $r_{\rm in}$ and an outer radius $r_{\rm out}$.  All the cores
are supposed to have an identical mass density $\rho=1$~g~cm$^{-3}$ so
that $R_i=[3m_i/(4 \pi \rho)]^{1/3}$.
%All the cores have initially the same mass in the range
%0.1--0.3~$\me$. 
We vary $r_{\rm in}$ and $r_{\rm out}$ while keeping
$r_{\rm out}-r_{\rm in}=1.5$--2~au in most of the simulations.  From
equation~(\ref{Mdust}), the mass of dust in between those two radii is
$\sim 5$--10~$\me$ initially.  Therefore, we will take $N \le 100$ or
$N \le 10$ when starting with a population of cores with masses
$m_p=0.1 \; \me$ or $1 \; \me$, respectively.  The upper values of $N$
correspond to all the dust mass being accreted into cores, which
probably leads to overestimating the number of cores or/and their
masses.
  
As mentioned above, the disc is assumed to be truncated at an inner
radius $r_{\rm cav}=0.05$~au.  At time $t=\tg$, a gap opens up between
$r_{10}=2$~au and $r_{20}=3$~au.  We take $\tg=3$~Myr, at which time
the disc mass has decreased by a factor of 10 (eq.~[\ref{sigmatgap}]
with $\eta=0.1$).  The cores start to migrate at a time $\tst$ which
we take to be either $1.5$, 2 or 2.5~Myr.  For
$\Sigma_1=2.7 \times 10^3 \; \gcm$, this corresponds to the surface
density at 1~au being reduced to $1.5 \times 10^3$, $10^3$ or
$680 \; \gcm$, respectively (see eq.~[\ref{sigmatgap}]), while for
$\Sigma_1=900 \; \gcm$, it corresponds to $495$, $360$ or
$225 \; \gcm$, respectively. Note that, for $t$ between 0 and $\tg$,
the important parameters are not $\tg$ and $\tst$ taken separately,
but $\tst / \tg$, as this is what determines the surface density in
the disc when the cores start to migrate.  Therefore, the same results
would be obtained with $\tg=1$~Myr and $\tst=0.5$, $0.7$ or 0.8~Myr.

For $\tnu$, we adopt values between $2 \times 10^4$ and
$6 \times 10^5$~years.

The simulations are run at least up to the time when
  there is no gas left in the disc in the regions were the planets
  are.  In some cases, the simulations are run for much longer, to
  check the stability of the systems over Myr timescales.

\subsection{Results}

\subsubsection{Starting with small cores in the inner disc}

\begin{figure}
\begin{center}
%\hspace{-1.cm}
\includegraphics[scale=0.33, angle=270]{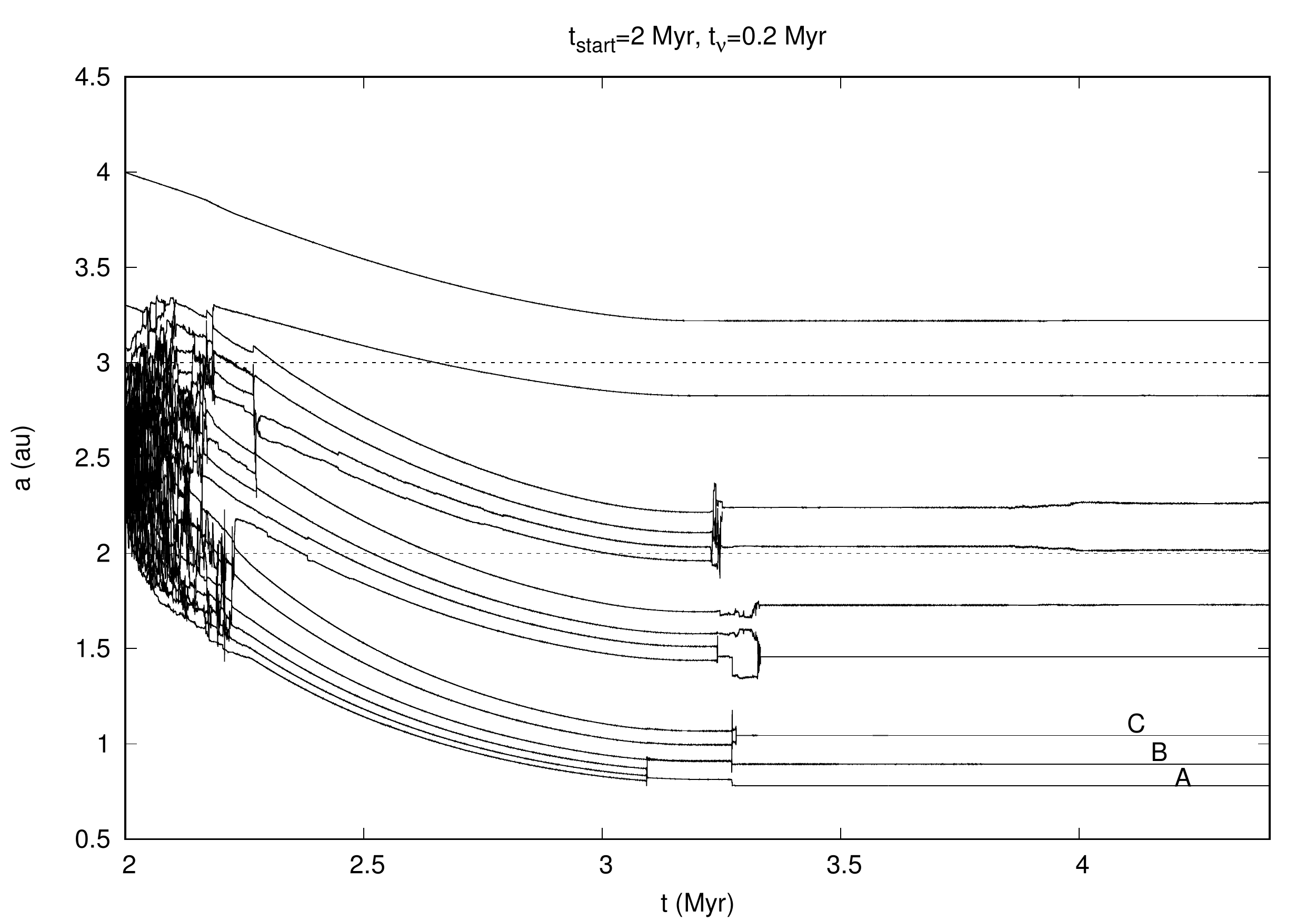}
\end{center}
%\vspace{-1.cm}
%%%%%% RUN7
\caption{Evolution of the semi--major axes (in units of au) of $N=50$
  cores in the system versus time (in units of Myr).  Here
  $\Sigma_1 = 2.7 \times 10^3 \; \gcm$, the cores are initially spread
  between $r_{\rm in}=2$~au and $r_{\rm out}=3.5$~au and start to
  migrate in at $\tst =2$~Myr.  The gap opens up at $\tg=3$~Myr between
  2 and 3~au ({\em dotted lines}), and subsequently the inner disc
  (within 2 au) dissipates on the timescale $\tnu=0.2$~Myr.
  Initially, all the cores have a mass $m_p=0.1 \; \me$.  The solid
  lines correspond to the different cores.  A line terminates just
  prior to a collision. The masses of the cores labelled 'A', 'B' and
  'C' are 0.5, 0.7 and 1.5~$\me$, respectively. The masses of the other
  cores left at 3.6~Myr are between 0.1 and 0.9~$\me$. }
\label{fig2}
\end{figure}

We start by calculating the evolution of a population of $N=50$ cores
with initial masses $m_p=0.1 \; \me$ and initially spread between 2
and 3.5~au.  Figure~(\ref{fig2}) shows the evolution of the
semi--major axes of these cores for
$\Sigma_1 = 2.7 \times 10^3 \; \gcm$, $\tst=2$~Myr and $\tnu=0.2$~Myr.
At $t=\tg = 3$~Myr, a gap opens up between 2 and 3~au.  At that point,
the planet with $a>3$~au continues to migrate in.  As it is in mean
motion resonances (MMR) with the planets in the gap, it pushes them
inwards.  Migration of planets in MMR is accompanied by an increase of
the eccentricities.  As there is no gas left in the gap to damp the
eccentricities, collisions among planets occur and resonances are
destroyed.  Similarly, the planets which are within 2~au continue to
migrate in, and as the gas dissipates collisions occur.  By the end of
the simulations, only the two outermost planets are still in MMR and
there is no gas left around the planets, which therefore do not
migrate anymore.  The three innermost planets, labelled 'A', 'B' and
'C', have $a=0.8, 0.9$ and 1~au and a mass of $0.5, 0.7$ and
1.5~$\me$, respectively.  The other cores left at the end of the
simulation have a mass between 0.1 and 0.9~$\me$.

The eccentricities of the planets left at the end of the simulation
are below 0.08 but they may increase on a longer timescale due to
gravitational interactions.  We note that the mutual
  spacing between the three innermost planets is very close to 12
  times their mutual Hill radii.  This has been shown by Pu \& Wu
  (2015) to be the limit for stability on Gyr timescales, with smaller
  separations leading to instabilities.  Therefore further collisions
  cannot be ruled out.  

\begin{figure}
\begin{center}
%\hspace{-1.cm}
\includegraphics[scale=0.4, angle=0]{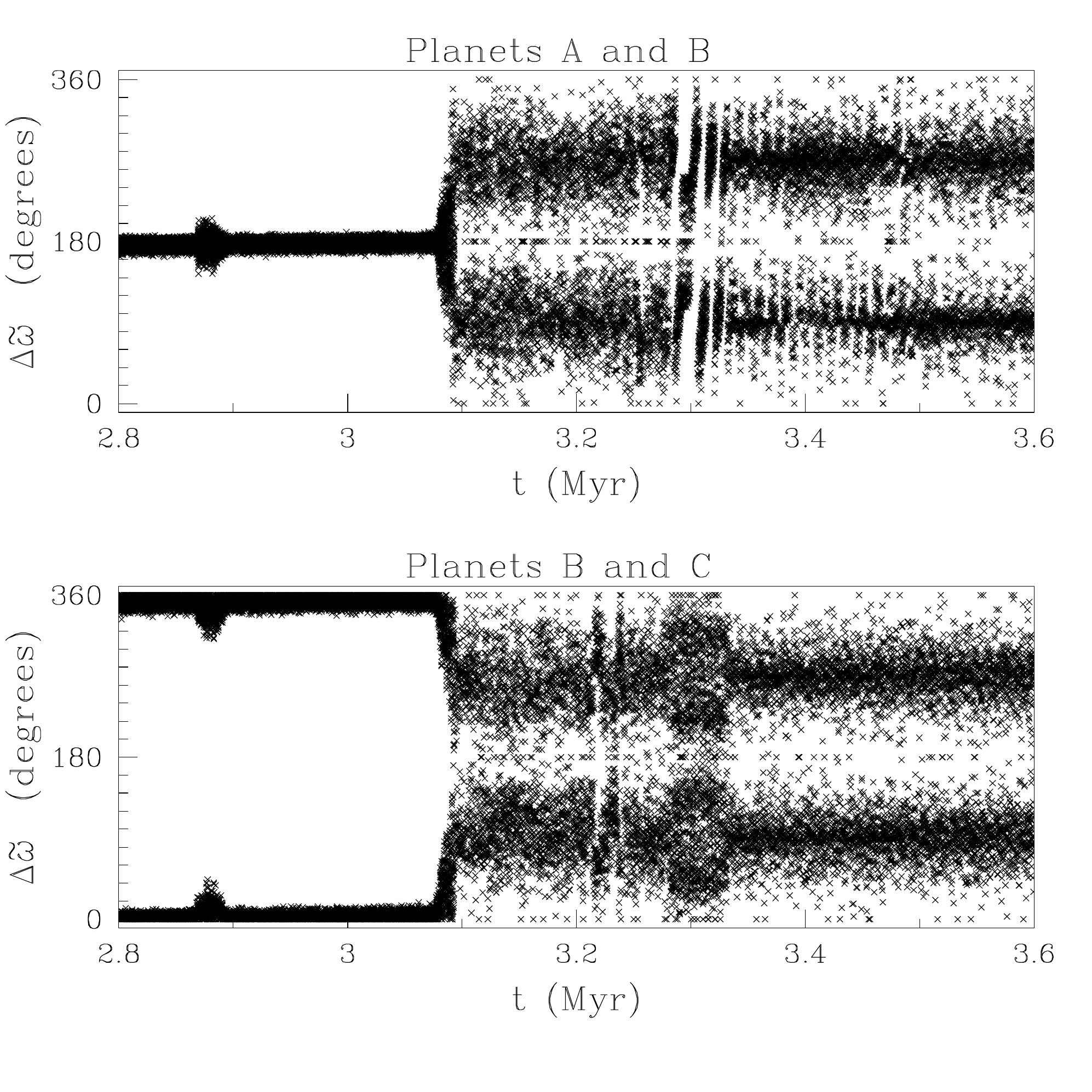}
\end{center}
%\vspace{-1.cm}
%%%%%% RUN7
\caption{Difference of the longitudes of pericentre $\Delta \varpi$
  (in degrees) for planets A and B ({\em upper plot}) and for planets B
  and C ({\em lower plot}) shown in figure~(\ref{fig2}) versus time
  (in units of Myr).  Before $t=3.1$~Myr, these planets are in MMR
  with $\Delta \varpi$ librating around either 180$^{\circ}$ or 0.  A
  collision during which planet A merges with another core occurs at
  $t \simeq 3.1$~Myr and destroys the resonances.  At the end of the
  simulation, there is no MMR among the three innermost planets.  }
\label{fig3}
\end{figure}

Figure~(\ref{fig3}) shows the difference of the longitudes of
pericentre $\Delta \varpi$ for planets A and B and for planets B and C
as a function of time.  Before $t=3.1$~Myr, these planets are in MMR.
For planets A and B, $\Delta \varpi$ librates around 180$^{\circ}$,
i.e.  the apsidal lines are anti--aligned and conjunction occurs when
one planet is near pericentre and the other near apocentre.  For
planets B and C, $\Delta \varpi$ librates around 0, i.e. the apsidal
lines are aligned and conjunction occurs when the planets are near
pericentre.  At $t \simeq 3.1$~Myr, a collision occurs that results in
planet A merging with another core, and this destroys the MMR.  At
$t \simeq 3.3$~Myr, planet C itself merges with another core.  At the
end of the simulation, there is no MMR among the three innermost
planets (although there is some degree of dynamical coupling).

The case illustrated in figures~(\ref{fig2}) and (\ref{fig3}) is very
typical.  All the simulations we have performed starting with
0.1~$\me$ cores and $\tst=2$~Myr and in which some of the cores were
able to migrate below 2~au before the gap opened up ended up with a
few planets between 0.5 and $\sim 1$~au.  The same
  simulation ran with a value of $\tnu$ ten times smaller,
  i.e. $\tnu=0.02$~Myr, gives qualitatively the same
  results.  Similar results were also obtained with $\tst=1.5$~Myr by
reducing the disc's mass (e.g., adopting $\Sigma_1=900 $ rather than
$2.7 \times 10^3 \; \gcm$).  The final masses are usually between a
fraction of an Earth mass and $\sim 1 \; \me$.  The fact that the
planets dot not migrate further down is not surprising as the
migration timescale for cores with masses between 0.1 and 1~$\me$ in a
disc with a surface mass density equal to a few $10^2 \; \gcm$ is
between 1 and 10~Myr.  For $\tst=2$~Myr, we obtained planets at the
inner edge of the disc (0.05~au) only when starting with cores more
massive than 0.3~$\me$.  By reducing $\tst$ to 1.5~Myr, we could
obtain planets near the edge when starting with cores more massive
than 0.15~$\me$.  In this case, MMR could be maintained.

In the simulation shown in figure~(\ref{fig2}), where the cores were
initially spread between $r_{\rm in}=2$~au and $r_{\rm out}=3.5$~au, 5
cores with a total mass of 4~$\me$ reached the inner disc (within
2~au).  The same simulation with $r_{\rm in}=3$~au and
$r_{\rm out}=5$~au ends up with only one 1~$\me$ core slightly below
2~au, all the others being left at larger distances from the star.  On
the other hand, by reducing $r_{\rm in}$ and $r_{\rm out}$ to 1.5 and
3~au, respectively, we obtain 8 cores with a total mass of 4.6~$\me$
between 0.5 and 2~au, the most massive core having a mass of 1~$\me$.
Again, the orbits of these cores are expected to evolve and further collisions
to occur on a timescale longer than that of the simulation.

When the cores do not migrate significantly, either because the
initial disc is not massive enough (low $\Sigma_1$) or migration
starts late, after the disc's mass has decreased down to low values
(large $\tst$), evolution is very much like what is obtained in {\em
  in situ} formation models.  There is not enough gas to damp the
eccentricities, so collisions occur and a few planets with masses at
most between 1 and 2~$\me$ form without significant inward migration. 

\subsubsection{Starting with heavier cores beyond the snow line}

\begin{figure}
\begin{center}
%\hspace{-1.cm}
\includegraphics[scale=0.33, angle=270]{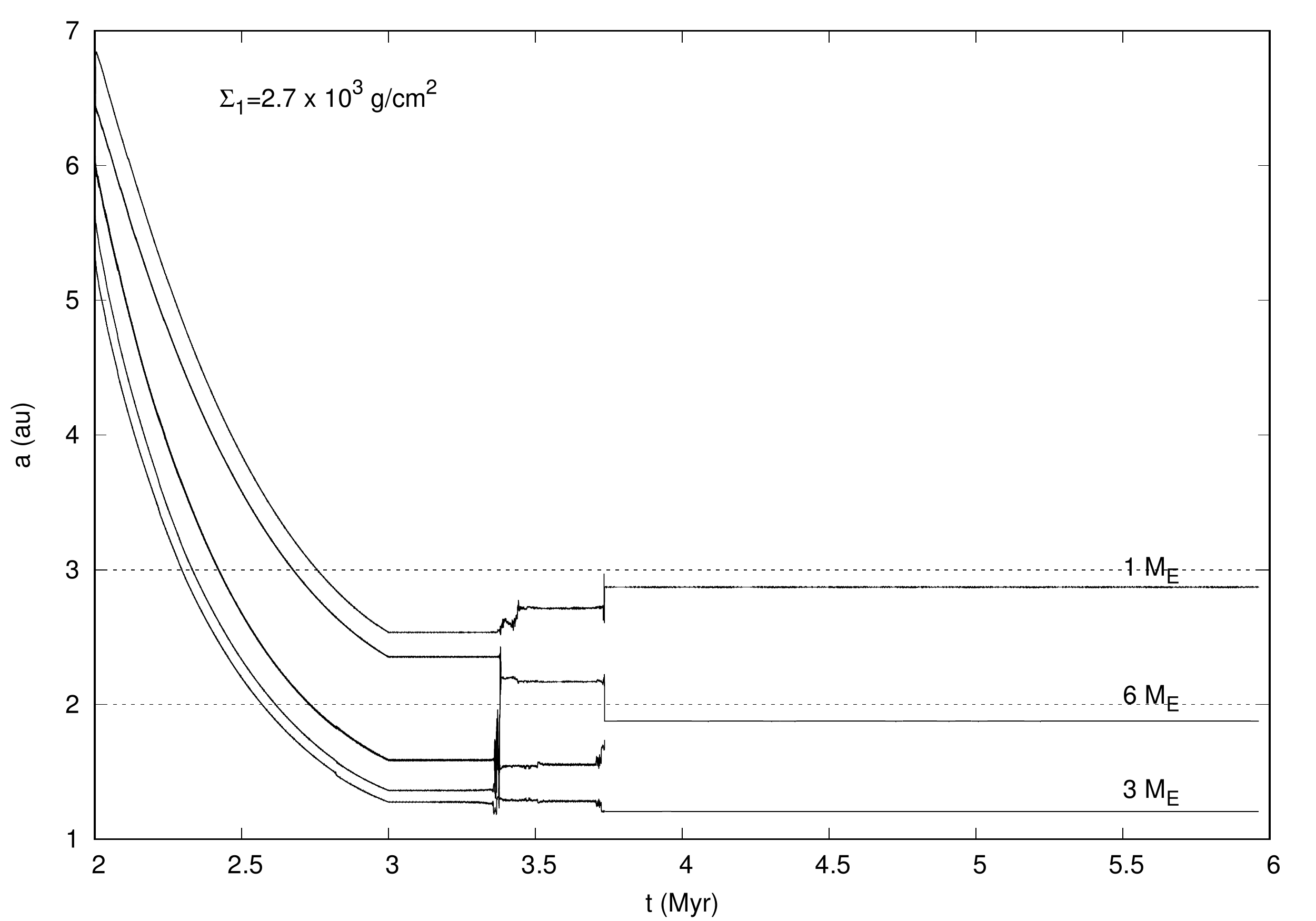}
\includegraphics[scale=0.33, angle=270]{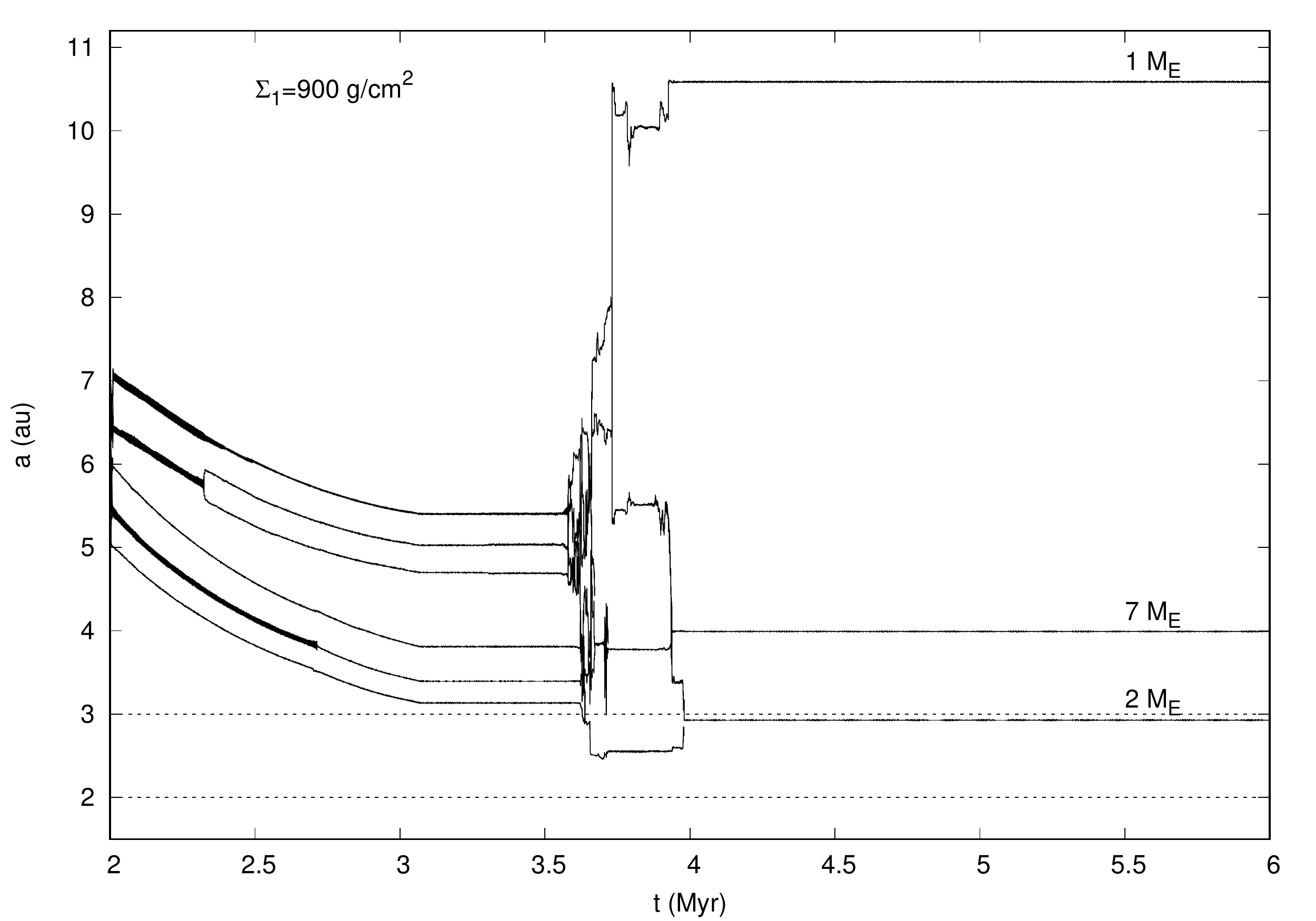}
\end{center}
%\vspace{-1.cm}
%%%%%% RUN31 and RUN35
\caption{Same as figure~(\ref{fig2}) but for $N=10$, $m_p=1 \; \me$,
  $r_{\rm in}=5$~au, $r_{\rm out}=6.5$~au and
  $\Sigma_1=2.7 \times 10^3 \; \gcm$ ({\em upper plot}) or
  $900 \; \gcm$ ({\em lower plot}).  Collisions between initial cores
  occur very early in the simulations so the 10 cores are not visible
  on the plots. Also heavy lines indicate several cores which
  oscillate around the same semi--major axis. 
The numbers above the lines at the end of the simulations indicate the
mass of the cores left in $\me$.}
\label{fig4}
\end{figure}

So far, we have ignored in the simulations heavier cores that may be
delivered to the inner parts from further away in the disc.  Whether
these cores will be present or not depends on wether they can form on
a timescale of 1~Myr.  Let us assume this the case, and there is a
population of 1~$\me$ cores that form within $\sim 1$~Myr beyond the
snow line at around 5~au (see Chambers 2016 for models).  According to
equation~(\ref{tmig}), the migration timescale of these cores at 5~au
would be the same as that of 0.1~$\me$ cores at 1~au, and therefore we
choose the same value of $\tst=2$~Myr as in the simulations with 
smaller cores presented above (we also keep $\tnu=0.2$~Myr).  In
figure~(\ref{fig4}), we show the evolution of 10 cores with  initial
mass of $m_p=1 \; \me$ and initially spread between 5 and 6.5~au, for
both $\Sigma_1=2.7 \times 10^3$ (upper plot) and
$900 \; \gcm$  (lower plot).  As pointed out above,
if these cores had formed {\em in situ} they would have used all the
dust initially present in the disc at these locations.  Alternatively,
they could have formed further away and migrated in, although it takes
more time to form cores at larger distances from the star.  As seen on
the figure, these cores grow through collisions and migrate in while
maintaining MMR.  When the gap opens up, in the case where
$\Sigma_1=2.7 \times 10^3 \; \gcm$, some of the cores have reached the
inner parts of the disc below 2~au while the other cores are in the
gap.
%Although the cores below 2~au could keep migrating as
%gas has not completely dissipated there, they stop because they are
%locked in MMRs with the cores in the gap, which themselves are not
%migrating anymore.  
The depletion of gas leads to eccentricity
growth and further collisions occur, leaving at the end of the
simulation two cores with masses of 3 and 6~$\me$ between 1 and 2~au
and one core with a mass of 1~$\me$ near 3~au.  In the case of a less
massive disc, i.e. $\Sigma_1 = 900 \; \gcm$, none of the cores reach
the gap before it opens up.  Here, by the end of the simulation, a
1~$\me$ core has been scattered to large distance ($\sim 11$~au)
while two cores with masses of 7 and 2~$\me$ are left at 4 and
$\sim 3$~au.  In that case, evolution is very much like what is
obtained with {\em in situ} models.   

As can be seen from the figure, the systems are stable
  after the last collisions occur and until 6 Myr.  The
  eccentricities, although large, vary smoothly and in a periodic
  way.  However, further collisions over much longer timescales cannot
  be ruled out. 

Cores with a mass of a few $\me$ could be obtained at smaller distances
from the star by starting the migration at earlier times.  

\subsubsection{Starting with a mixture of small and heavier cores}

If we now start with both small cores ($m_p = 0.1 \; \me$) at around
2~au and heavier cores ($m_p = 1 \; \me$) at around 5~au, the
evolution can be predicted on the basis of the results presented
above.  The small cores evolve as described above, and always end up
forming planets with masses at most between 1 and 2 $\me$ at locations
that depend on how fast they have migrated, but usually between 0.5
and 2~au.  If the heavier cores are able to catch up, they may sweep
some of the smallest ones on their way in and the final system has a
few cores of several earth masses below 2~au, mixed with less massive
planets.  On the other hand, the heavier cores may not be able to
migrate below the inner edge of the gap (e.g., for low values of
$\Sigma_1$ or if the heavy cores have large $\tst$), in which case
they will also evolve mostly {\em in situ} to form a few massive cores
beyond 4~au or so.

\begin{figure}
\begin{center}
%\hspace{-1.cm}
\includegraphics[scale=0.33, angle=270]{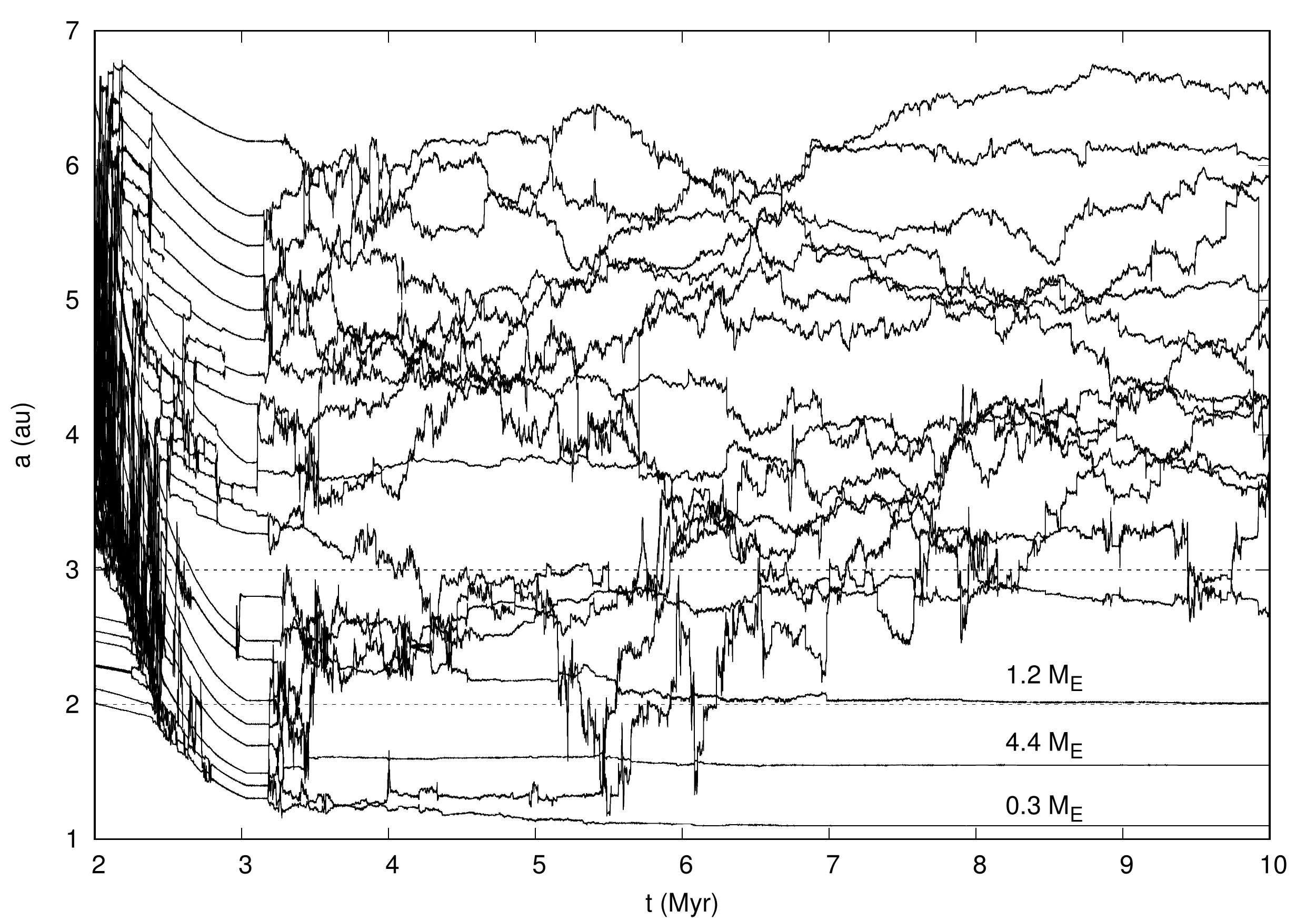}
\end{center}
%\vspace{-1.cm}
%%%%%% RUN33
\caption{Same as figure~(\ref{fig2}) but here there are initially 50
  cores with  $m_p=0.1 \; \me$ between 2 and 5~au and 5 cores with
  $m_p=1 \; \me$ between 5 and 6.5~au.  The numbers above the lines at
  the end of the simulation indicate the mass of the cores left in
  $\me$.}
\label{fig5}
\end{figure}

This is exactly what we have found in simulations starting with 50
small cores ($m_p = 0.1 \; \me$) between 2 and 5~au together with 5 or
10 heavier cores ($m_p = 1 \; \me$) between 5 and 6.5~au.  We
considered $\Sigma_1 = 2.7 \times 10^3$ or $900 \; \gcm$ and, as
above, $\tst=2$~Myr and $\tnu=0.2$~Myr.  The case
  $\Sigma_1 = 2.7 \times 10^3 \; \gcm$ is illustrated in
  figure~(\ref{fig5}).  The simulation has been stopped after 8~Myr
  but here of course more collisions are going to occur on longer
  timescales.  The same simulation with $\Sigma_1 =900 \; \gcm$ does
  not show any significant migration.  In that case, the cores evolve
  mostly {\em in situ}, and after the disc dissipates there is no core
  below 1.8~au.

\section{Summary and discussion}
\label{sec:discussion}

In this paper, we have studied the outcome of core migration in
transition discs produced by photoevaporation.  The initial surface
mass density in discs at around 1~au is at most on the order
$10^3 \; \gcm$, and it decreases with time.  Therefore, planetary
cores that form at this distance from the star migrate on a timescale
smaller than the disc lifetime only if their mass is larger than about
0.1~$\me$.  As it takes at least about 1~Myr to form such cores at
around 1~au, migration of cores forming below the snow line starts to
be relevant only when the disc is at least 1~Myr old.  Similarly, if
the surface density varies as $r^{-1}$, cores at $\sim 5$~au have a
migration timescale shorter than the disc lifetime only when their
mass reaches about 1~$\me$. Assuming these cores form on a timescale
of about 1~Myr, they will also start to migrate significantly when the
disc is about 1~Myr old.

In low~mm flux transition discs, a gap is believed to open up at around
2--3~au due to X--ray photoevaporation when the disc is a few Myr.  The
disc's inner parts are subsequently accreted onto the central star as a
result of viscous evolution, while the outer edge of the gap recedes
under the effect of photoevaporation.  In this context, we find that
$\sim 0.1 \; \me$ cores that form within $\sim 1$~Myr between 1--4~au
end up forming a system of a few planets with masses
between a fraction of an Earth mass and 1.5~$\me$ at most.  In
general, these cores do not migrate down further than $\sim 0.5 $~au,
and may even evolve {\em in situ} in low mass discs.   They are not in mean
motion resonances.  Such resonances, which appear during the migration
phase, are usually destroyed when the gas dissipates (as previously
found by Cossou et al. 2014 and Coleman \& Nelson 2016).  
It is difficult to form planets at smaller radii, unless we start with
heavier cores or start migrating the cores earlier (as, e.g., in
Terquem \& Papaloizou~2007).  In both cases, it
would require a more efficient planet formation process than envisioned
here.  
If $\sim 1$~$\me$ cores can form beyond
the snow line also within $\sim 1$~Myr, they migrate on the same
timescale as the smaller cores and, depending on the surface density
in the disc, they may or may not reach the inner parts of the disc.
After the gap has opened up, the inner parts of the disc become
isolated from the outer parts, and no more material is delivered
there. 

We have found that, in a disc with an initial surface mass density of
about $10^3 \; \gcm$ at 1~au, it was possible to obtain a system of
cores similar to that of the Solar system, with a few planets with
masses of a few tenths of an earth mass to an earth mass in the
terrestrial zone and a more massive core a little bit further away
(fig.~[\ref{fig2}] and lower panel of fig.~[\ref{fig4}]).  However, as
the photoevaporation model predicts that the outer edge of the gap is
receding rather fast, it is difficult to envision within this model
how the massive core could accrete a gaseous envelope to become a
giant planet.  Such a system of low mass planets is
  similar to those obtained by Cossou et al. (2014) and Coleman \&
  Nelson (2016) in their models with inefficient migration. 

The exact parameters that should be used in such simulations are of
course not known, and different outcomes could be produced by changing
them.  However, the point of this paper is to show that, by adopting
reasonable parameters that fit the observations and the current
understanding of planet formation and migration, systems of
cores/planets can be obtained where all the planets do not end up on
resonant chains near the disc's inner edge and delivery of material to
the inner parts from the outer parts of the disc can be avoided.  To
that extent, the type of systems that we obtain resembles more our
Solar system than the systems observed by {\em Kepler}, which is
biased towards detecting planets on very tight orbits.  Our results
predict that low mm flux transition disc may harbour terrestrial
planets in the habitable zone.  

Note that, as collisions between cores occur until
  rather late in the disc's evolution, dust may be produced at
  distances below 1 or 2~au from the central star even after the
  disc's inners parts have been accreted.  In that case, the disc
  would not appear as transition disc.  

This study suggests that low mm flux transition discs may not be able to
form giant planets before X--ray photoevaporation opens up a gap.
These discs would then form planetary systems of the type obtained in
this study: predominantly low mass planets at around 1~au with
possibly more massive cores further away.  High mm flux transition
discs, by contrast, are associated with more massive stars and may
form giant planets of a few Jupiter masses, massive enough to open up
a gap, before photoevaporation could proceed.

%\section*{Acknowledgements}

%
%===========================================================
%

%\bibliographystyle{apj}
%\bibliographystyle{plain}
%\bibliographystyle{mn2e}
%\bibliography{biblio_papers}

%
%===========================================================

\label{lastpage}
\end{document}